\documentclass[twocolumn]{aastex63}

\newcommand\aastex{AAS\TeX}
\newcommand\latex{La\TeX}

\received{October 13, 2020}

\shorttitle{GC Luminosity Function of NGC1052-DF2 and DF4}
\shortauthors{Shen et al.}
\graphicspath{{./}{figures/}}

\begin{document}

\title{A Complex Luminosity Function for the Anomalous Globular Clusters in NGC1052-DF2 and NGC1052-DF4}


\correspondingauthor{Zili Shen}
\email{zili.shen@yale.edu}

\author[0000-0002-5120-1684]{Zili Shen}
\affiliation{Astronomy Department, Yale University,
52 Hillhouse Ave,
New Haven, CT 06511, USA}

\author{Pieter van Dokkum}
\affiliation{Astronomy Department, Yale University,
52 Hillhouse Ave,
New Haven, CT 06511, USA}

\author{Shany Danieli}
\altaffiliation{NASA Hubble Fellow}
\affiliation{Astronomy Department, Yale University,
52 Hillhouse Ave,
New Haven, CT 06511, USA}
\affiliation{Department of Physics, Yale University, New Haven, CT 06520, USA}
\affiliation{Yale Center for Astronomy and Astrophysics, Yale University, New Haven, CT 06511, USA}
\affiliation{Institute for Advanced Study, 1 Einstein Drive, 
Princeton, NJ 08540, USA}

\begin{abstract}

NGC1052-DF2 and NGC1052-DF4 are ultra-diffuse galaxies (UDGs) that were found to have extremely low velocity dispersions, indicating that they have little or no dark matter. Both galaxies host anomalously luminous globular cluster (GC) systems, with a peak magnitude of their GC luminosity function (GCLF) that is $\sim1.5$ magnitudes brighter than the near-universal value of $M_V \approx -7.5$.
Here we present an analysis of the joint GCLF of the two galaxies, making use of new HST photometry and Keck spectroscopy, and a recently improved distance measurement. We apply a homogeneous photometric selection method to the combined GC sample of DF2 and DF4.
The new analysis shows that the peak of the combined GC luminosity function remains at $M_V \approx -9$ mag. In addition, we find a subpopulation of less luminous GCs at $M_V \approx -7.5$ mag, where the near-universal GCLF peak is located.  
The number of GCs in the magnitude range of $-5$ to $-8$ is $7.1_{-4.34}^{+7.33}$ in DF2 and $8.6_{-4.83}^{+7.74}$ in DF4,
similar to that expected from other galaxies of the same luminosity.
The total GC number between $M_V$ of $-5$ to $-11$ is $18.5_{-4.42}^{+8.99}$ for DF2 and $18.6_{-4.92}^{+9.37}$ for DF4 , calculated from the background-subtracted GCLF.
The updated total number of GCs in both galaxies is $37 ^{+11.08}_{-6.54}$. The number of GCs do not scale with the halo mass in either DF2 or DF4, suggesting that  $N_{GC}$ is not directly determined by the merging of halos.

\end{abstract}

\keywords{galaxies: star clusters --- 
galaxies: structure --- galaxies: individual(NGC1052-DF2, NGC1052-DF4)}

\section{Introduction} \label{sec:intro}
NGC1052-DF2 and NGC1052-DF4 (hereafter DF2 and DF4) are two ultra-diffuse galaxies that have similar size, color, and luminosity, and also share more unusual properties. Both have been found to have extremely low velocity dispersion, of their globular cluster systems as well as the diffuse light, and are thus thought to be dark matter deficient \citep{VanDokkum2018Nature,VanDokkum2019,Danieli2019_KCWI,Emsellem2019}. 
Moreover, both galaxies host a spectacular population of globular clusters (GCs). The GCs contribute 4\% of the total light, which is a factor of 100 more than the typical fraction \citep{VanDokkum2018enigma}. This large contribution comes not only from the high number of clusters but also from their unusual luminosities. Their globular cluster luminosity function (GCLF) show a peak around $M_V \approx -9$ mag, two magnitudes brighter than expected \citep{VanDokkum2018enigma,VanDokkum2019}.

It has been generally observed that the GCLF peaks at a near-constant absolute magnitude of around -7.5 mag \citep{Harris2001,Rejkuba2012GCLF}. Both in the Milky Way and beyond, GCLFs have been found to be well-fitted by a Gaussian distribution with the center at this absolute magnitude. \citet{Richtler2003} explored using this near-universal turnover luminosity as a distance indicator. Assuming a constant mass-to-light ratio, the observed Gaussian GCLF (measured in magnitudes) corresponds to a lognormal globular cluster mass function (GCMF). 
Dynamical evolution of GCs, resulting from evaporation, tidal shocks, and dynamical friction, can transform an initial power-law mass function into the observed lognormal GCMF \citep{Gnedin_Ostriker1997, Fall_Zhang2001, Vesperini2003, Kruijssen2009, Elmegreen2010}. Two-body evaporation and tidal shocks preferentially affect low mass and loosely-bound GCs, while dynamical friction disrupts the highest mass GCs and the ones closest to the galactic center \citep{Elmegreen_Hunter2010, Gieles_Renaud2016, Pfeffer2018, Li_Gnedin2019}. These effects act during different epochs of GC evolution. Early rapid tidal shocks from the gas-rich disk can create a single-peaked distribution from the power-law, and after GCs are dispersed into the halo, slow disruption by evaporation lowers the peak to the observed value \citep[see, e.g.][]{Kruijssen2015}.

Against the background of a near-universal GCLF, the apparent offset in the DF2/DF4 GCLF peak magnitude led to skepticism regarding the distance measurement to DF2 and DF4.  \citet{Trujillo2019} and \citet{Monelli2019} respectively proposed distances of 13 Mpc to DF2 and $14.2 \pm 0.7$ Mpc to DF4. Compared to the 19 Mpc original estimate of \citet{vanDokkum2018dist}, this decrease in distance would bring the GCLF peak in agreement with other galaxies, but it would not resolve the high fraction of light contained in the GCs. 

Later, \citet{Danieli2020} measured the Tip of the Red Giant Branch (TRGB) distance from deep \textit{Hubble space telescope} (HST) data and confirmed that DF4 is at a distance of $20.0 \pm 1.6$ Mpc. Thus, an erroneous distance cannot explain why the peak of their GCLF is offset from the canonical $M_V \approx -7.5$.

The unusual GCLF in DF2 and DF4 is difficult to interpret with existing data. One possibility is that the luminosities of all GCs in DF2 and DF4 are systematically higher. The other interpretation is that in addition to normal GCs, there is a subpopulation of high luminosity GCs and only the luminous subpopulation has spectroscopic confirmation. With the small numbers of confirmed GCs in each galaxy (twelve in DF2, seven in DF4), the data are insufficient to distinguish between the two. Furthermore, if the normal subpopulation exists, it would be fainter than the spectroscopic completeness limit (for GC candidates in DF2 is F814W$ < 23$ which corresponds to $M_V \approx -8.5$).

\begin{figure*}[ht!]
\centering
\includegraphics[width=0.75\textwidth]{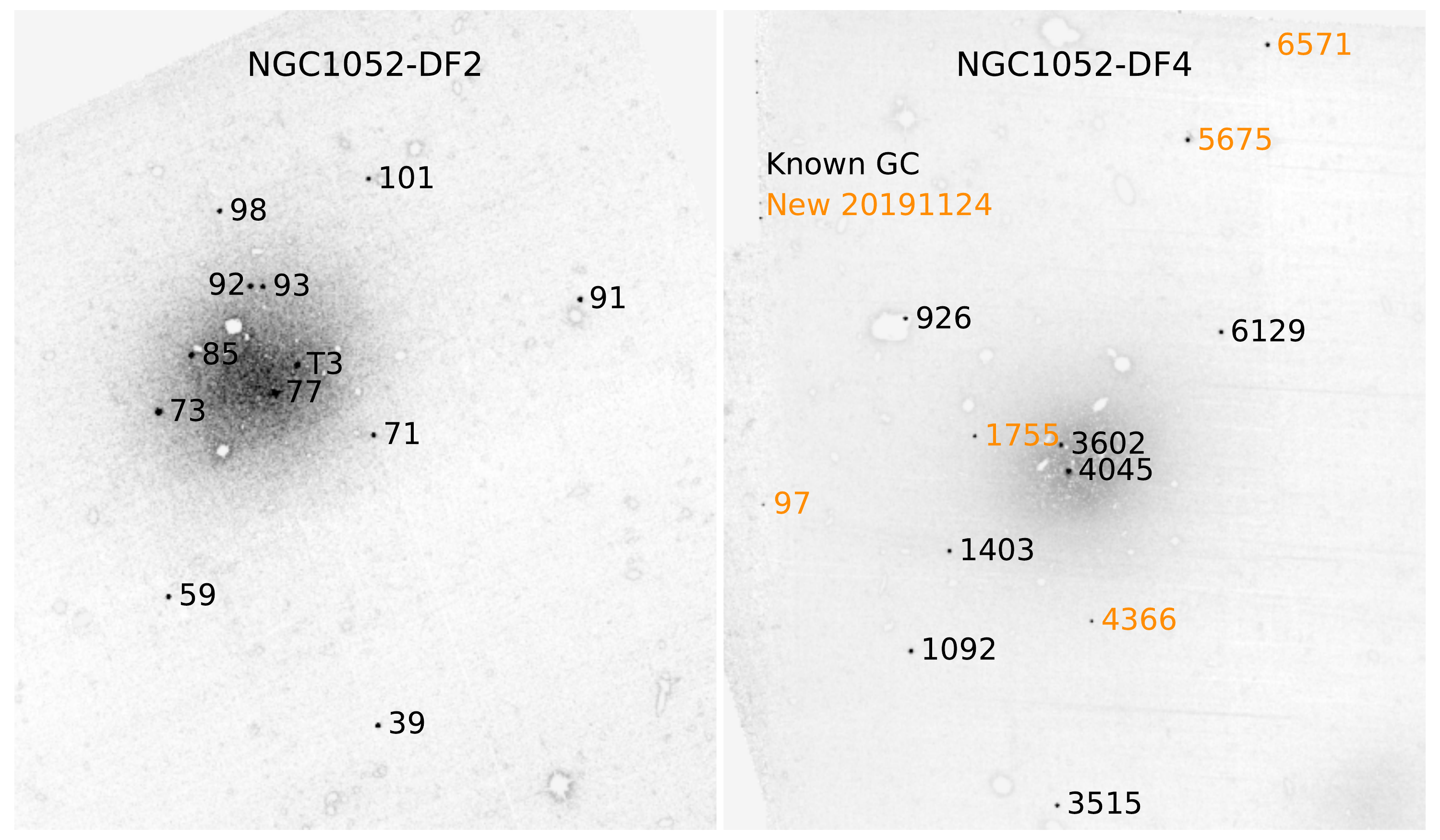}
\caption{\textit{HST} ACS F814W image of NGC1052-DF2 and NGC1052-DF4, with all compact sources masked out except for spectroscopically confirmed GCs labelled by their ID numbers. The mask is created from \texttt{SourceExtractor} segmentation map. Due to new and deeper photometry, the ID number used in this work is different from that of \citet{VanDokkum2019} and the cross-identification are listed in Table \ref{tab:DF4}.  }\label{fig:intro}
\end{figure*}

Here we present a combined analysis of the GCLF in DF2 and DF4, made possible by new data from HST with an expanded footprint \citep[see][]{Danieli2020} and new spectroscopy of candidate GCs. The HST images used for this analysis are shown in Figure \ref{fig:intro}. We analyze the combined GC sample in the two galaxies in a uniform manner, and update the photometric selection procedure.

\section{Spectroscopy} \label{sec:spec}

\begin{figure}[ht!]
\centering
\includegraphics[width=0.9\columnwidth]{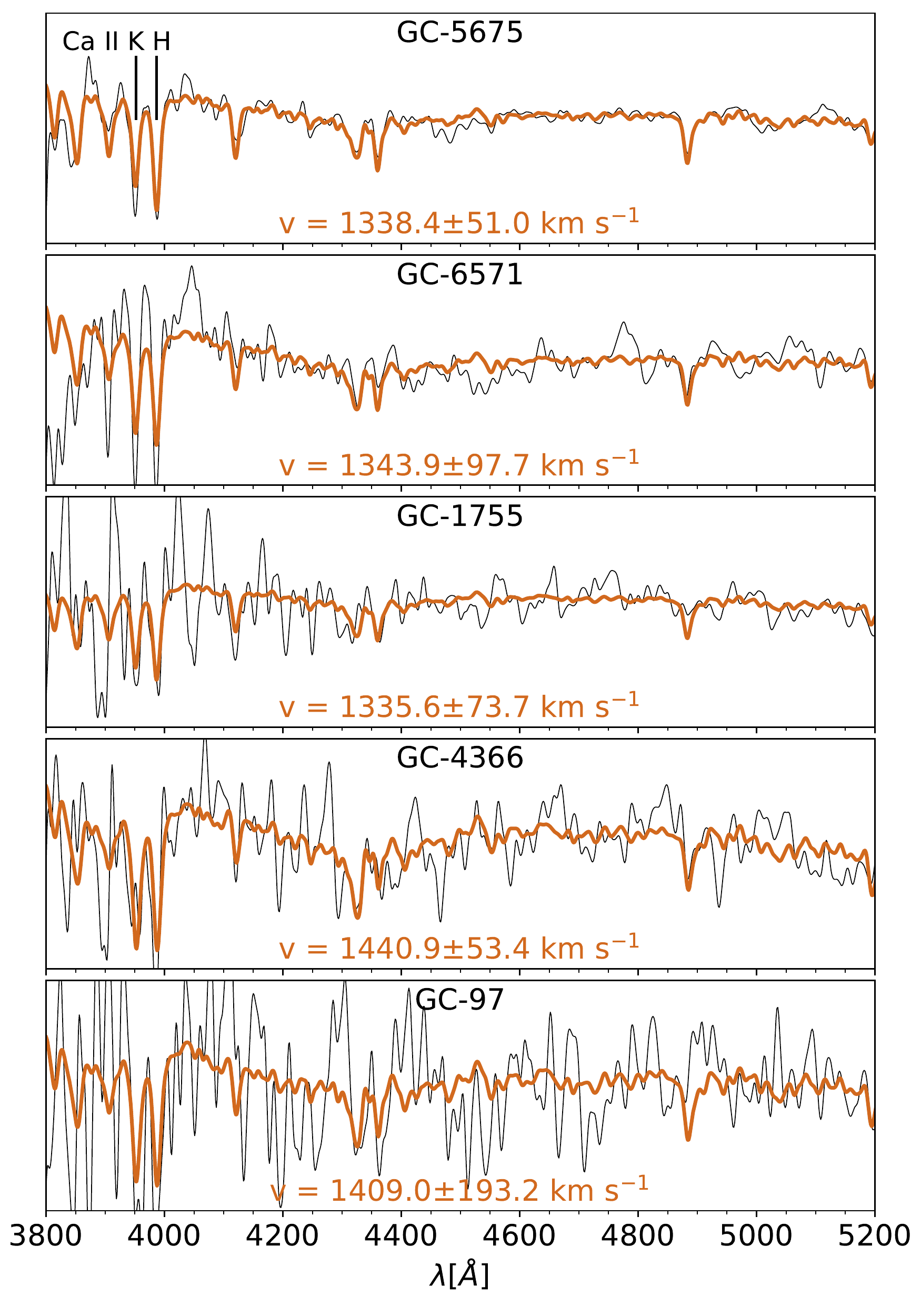}
\caption{Keck/LRIS spectra of five newly confirmed GCs in NGC1052-DF4, overplotted with the best-fit model and heliocentric radial velocity from \texttt{pPXF}. \label{fig:spec}}
\end{figure}

\begin{deluxetable*}{ccccccccc}[ht!]
\tablecaption{Spectroscopically Examined Objects in NGC 1052-DF2. \label{tab:DF2}}
\tabletypesize{\footnotesize}
\tablehead{\colhead{Id} & \colhead{Other ID \tablenotemark{a}}&\colhead{RA} & \colhead{Decl.}
 & \colhead{$M_{V,606}$} & \colhead{$V - I$}& \colhead{$v$ \tablenotemark{b}} \\
 & & (J2000) & (J2000) & [mag]  & & [km\,s$^{-1}$]}
\startdata
NGC1052-DF2 & [KKS2000]04 & 2$^{\rm h}41^{\rm m}46.8^{\rm s}$&
$-8\arcdeg24\arcmin09\farcs3$ & $-15.3$& 0.40 & $1805 \pm 1.1$\\
GC-39 & & 2$^{\rm h}41^{\rm m}45.07^{\rm s}$ & $-8\arcdeg25\arcmin24\farcs9$
& $-9.3$& 0.36& $1818_{-7}^{+7}$ \\
GC-59 & &2$^{\rm h}41^{\rm m}48.08^{\rm s}$ & $-8\arcdeg24\arcmin57\farcs5$
& $-8.9$& 0.34&    $1799_{-15}^{+16}$ \\
GC-71 & & 2$^{\rm h}41^{\rm m}45.13^{\rm s}$ & $-8\arcdeg24\arcmin23\farcs0$
& $-9.0$ & 0.38 &$1805_{-8}^{+6}$ \\
GC-73 & & 2$^{\rm h}41^{\rm m}48.22^{\rm s}$ & $-8\arcdeg24\arcmin18\farcs1$
& $-10.1$& 0.32 &$1814_{-3}^{+3}$ \\
GC-77 & & 2$^{\rm h}41^{\rm m}46.54^{\rm s}$ & $-8\arcdeg24\arcmin14\farcs0$
& $-9.6$& 0.36&$1804_{-6}^{+6}$  \\
GC-80 & T3 & 2$^{\rm h}41^{\rm m}45.21^{\rm s}$ & $-8\arcdeg23\arcmin28\farcs3$
& $-8.6$& 0.33 &$1794.1_{-6.4-10.1}^{+5.9+5.1}$  \\
GC-85 & & 2$^{\rm h}41^{\rm m}47.75^{\rm s}$ & $-8\arcdeg24\arcmin05\farcs9$
& $-9.2$ & 0.39 &$1801_{-6}^{+5}$ \\
GC-91 &&  2$^{\rm h}41^{\rm m}42.17^{\rm s}$ & $-8\arcdeg23\arcmin54\farcs0$
& $-9.2$ & 0.36& $1802_{-10}^{+10}$\\
GC-93 & &2$^{\rm h}41^{\rm m}46.72^{\rm s}$ & $-8\arcdeg23\arcmin51\farcs3$
& $-8.6$ & 0.39&$1818.6_{-6.9-4.4}^{+5.5+1.1}$  \\
GC-92 &  &2$^{\rm h}41^{\rm m}46.90^{\rm s}$ & $-8\arcdeg23\arcmin51\farcs1$
& $-9.4$ & 0.40& $1789_{-7}^{+6}$ \\
GC-98 && 2$^{\rm h}41^{\rm m}47.34^{\rm s}$ & $-8\arcdeg23\arcmin35\farcs2$
& $-8.7$& 0.44& $1784_{-10}^{+10}$  \\
GC-101 && 2$^{\rm h}41^{\rm m}45.21^{\rm s}$ & $-8\arcdeg23\arcmin28\farcs3$
& $-8.6$& 0.45& $1800_{-14}^{+13}$  \\
\hline
&PN1  &  2$^{\rm h}41^{\rm m}47.42^{\rm s}$ & $-8\arcdeg23\arcmin41\farcs0$ & -- & -- &$1790.8_{-1.8}^{+2.0}$  \\
&PN2 & 2$^{\rm h}41^{\rm m}45.99^{\rm s}$ & $-8\arcdeg23\arcmin47\farcs0$ & -- & --& $1786.4_{-5.9}^{+7.2}$  \\
&PN3 & 2$^{\rm h}41^{\rm m}46.33^{\rm s}$ & $-8\arcdeg24\arcmin28\farcs2$ & --& -- & $1766.4_{-7.1}^{+10.2}$  \\
\hline
BG-21& & 2$^{\rm h}41^{\rm m}49.76^{\rm s}$ & $-8\arcdeg26\arcmin12\farcs5$  
& $-9.65$& 0.45& --  \\
BG-68& & 2$^{\rm h}41^{\rm m}45.58^{\rm s}$ & $-8\arcdeg24\arcmin33\farcs8$  & -10.62 &0.18&--  \\
Star &T7&  2$^{\rm h}41^{\rm m}47.35^{\rm s}$ & $-8\arcdeg23\arcmin45\farcs9$  & -9.65 &--& -- \\
\enddata

\tablenotetext{a}{T3 and T7 are proposed GC candidates in \citet{Trujillo2019} and are observed by \citet{Emsellem2019}. T3 is a GC in DF4 while T7 is a foreground star.}
\tablenotetext{b}{DF2 systemic velocity from \citet{Danieli2019_KCWI}; velocities of GC-80, GC-93, and all PNs from \citet{Emsellem2019}.}
\tablecomments{Grouped from top to bottom: the previously identified GCs with their alternative IDs; planetary nebulae in DF2; known background galaxies and foreground star contaminants. Within each group, the objects are ordered by ID number. The prefix in the first column denotes the object type: GC = globular cluster in DF4, BG = background galaxy, PN = planetary nebulae.}
\end{deluxetable*}

\begin{deluxetable*}{cccchccc}[ht!]
\tablecaption{Spectroscopically Examined Objects in NGC 1052-DF4\label{tab:DF4}. }
\tabletypesize{\footnotesize}
\tablehead{\colhead{Id} & \colhead{Other ID\tablenotemark{a}}&\colhead{RA} & \colhead{Decl.}  
& R&$M_{V,606}$ & $V-I$ \tablenotemark{b}  & $v$  \\
& & (J2000)  & (J2000) &(kpc)& [mag]  & [mag]& [km\,s$^{-1}$] }
\startdata
NGC1052-DF4 & --& 2$^{\rm h}39^{\rm m}15.11^{\rm s}$&
$-8\arcdeg6\arcmin58\farcs6$ & -- & -15.0&0.32&$1444.6^{+7.8}_{-7.7}$\\
GC-5675 &  --& 2$^{\rm h}39^{\rm m}12.07^{\rm s}$ & $-8\arcdeg5\arcmin59\farcs6$ & 7.20 & -9.43 & 0.34 & $1351.0_{-51.0}^{+51.0}$\\
GC-6571 & --& 2$^{\rm h}39^{\rm m}10.54^{\rm s}$ & $-8\arcdeg5\arcmin46\farcs0$ & 9.63 & -8.75 & 0.54 & $1357.0_{-98.0}^{+98.0}$\\
GC-1755 &  --& 2$^{\rm h}39^{\rm m}16.35^{\rm s}$ & $-8\arcdeg6\arcmin44\farcs4$ & 2.26 & -7.88 & 0.43 & $1348.0_{-74.0}^{+74.0}$\\
GC-4366 &  --& 2$^{\rm h}39^{\rm m}15.63^{\rm s}$ & $-8\arcdeg7\arcmin29\farcs8$ & 3.12 & -7.44 & 0.41 & $1454.0_{-53.0}^{+53.0}$\\
GC-97 & --& 2$^{\rm h}39^{\rm m}19.54^{\rm s}$ & $-8\arcdeg6\arcmin43\farcs5$ & 6.56 & -6.70 & 0.44 & $1422.0_{-190.0}^{+190.0}$\\
GC-926 & 2726 & 2$^{\rm h}39^{\rm m}16.74^{\rm s}$ & $-8\arcdeg6\arcmin15\farcs9$ & 4.76 & -9.19 & 0.42 & $1441.2_{-4.8}^{+4.9}$\\
GC-6129 & 2537 & 2$^{\rm h}39^{\rm m}12.52^{\rm s}$ & $-8\arcdeg6\arcmin40\farcs6$ & 4.10 & -9.14 & 0.41 & $1451.0_{-3.3}^{+3.6}$\\
GC-3602 & 2239 & 2$^{\rm h}39^{\rm m}15.22^{\rm s}$ & $-8\arcdeg6\arcmin52\farcs2$ & 0.65 & -8.60 & 0.41 & $1457.1_{-5.5}^{+4.6}$\\
GC-4045 & 1968 & 2$^{\rm h}39^{\rm m}15.24^{\rm s}$ & $-8\arcdeg6\arcmin58\farcs0$ & 0.21 & -9.70 & 0.40 & $1445.4_{-2.3}^{+2.6}$\\
GC-1403 & 1790 & 2$^{\rm h}39^{\rm m}17.24^{\rm s}$ & $-8\arcdeg7\arcmin5\farcs8$ & 3.15 & -8.98 & 0.33 & $1438.4_{-4.6}^{+4.8}$\\
GC-1092 & 1452 & 2$^{\rm h}39^{\rm m}18.23^{\rm s}$ & $-8\arcdeg7\arcmin23\farcs3$ & 5.10 & -9.13 & 0.39 & $1445.5_{-4.1}^{+4.0}$\\
GC-3515 & 943 & 2$^{\rm h}39^{\rm m}16.97^{\rm s}$ & $-8\arcdeg8\arcmin4\farcs5$ & 6.94 & -8.59 & 0.42 & $1445.1_{-5.2}^{+5.0}$\\
\hline
BG-19 & 2844 & 2$^{\rm h}39^{\rm m}20.37^{\rm s}$ & $-8\arcdeg6\arcmin17\farcs8$ & 8.55 & -8.88 & 0.46 & --\\
BG-8251 & 254 & 2$^{\rm h}39^{\rm m}11.54^{\rm s}$ & $-8\arcdeg9\arcmin2\farcs2$ & 13.04 & -7.58 & 0.34 &--\\
ST-5207 &  -- & 2$^{\rm h}39^{\rm m}14.40^{\rm s}$ & $-8\arcdeg6\arcmin59\farcs9$ & 1.02 & -9.69 & 0.35 & --\\
BG-6444 &  --& 2$^{\rm h}39^{\rm m}14.52^{\rm s}$ & $-8\arcdeg8\arcmin27\farcs7$ & 8.68 & -7.73 & 0.37 &--\\
BG-6180 &  --& 2$^{\rm h}39^{\rm m}14.29^{\rm s}$ & $-8\arcdeg8\arcmin4\farcs9$ & 6.53 & -7.59 & 0.11 &--\\
BG-4951 &  --& 2$^{\rm h}39^{\rm m}14.41^{\rm s}$ & $-8\arcdeg6\arcmin54\farcs3$ & 1.08 & -7.52 & 0.35 &--\\
BG-1173 &  --& 2$^{\rm h}39^{\rm m}18.48^{\rm s}$ & $-8\arcdeg7\arcmin49\farcs4$ & 6.93 & -7.45 & 0.39 &--\\
BG-367 &  --& 2$^{\rm h}39^{\rm m}16.97^{\rm s}$ & $-8\arcdeg5\arcmin31\farcs6$ & 8.86 & -7.00 & 0.45 & --\\
BG-6561 &  --& 2$^{\rm h}39^{\rm m}13.39^{\rm s}$ & $-8\arcdeg7\arcmin42\farcs7$ & 4.94 & -6.98 & 0.45 & --\\
\enddata
\tablenotetext{a}{Name from \citet{VanDokkum2019}. This entry is left empty for objects that are identified in this work and have no previous name.}
\tablenotetext{b}{$F606W-F814W$ from HST/ACS, in the AB system.}
\tablecomments{Grouped from top to bottom: the new GCs presented in this work; the previously identified GCs with their alternative IDs; known background galaxies and foreground star contaminants. Within each group, the objects are ordered by magnitude. The prefix in the first column denotes the object type: GC = globular cluster in DF4, BG = background galaxy, ST = foreground star.}
\end{deluxetable*}

\subsection{DF2}
In DF2, eleven GCs were previously identified with LRIS spectroscopy by \citet{VanDokkum2018Nature}. 
Eight other GC candidates were proposed by \citet{Trujillo2019} and VLT/MUSE spectroscopy was obtained by \citet{Emsellem2019}. Out of the eight candidates, one (T3) was confirmed to be associated with DF2 and one (T7) is a foreground star. In addition, \citet{Emsellem2019} also identified three planetary nebulae in DF2 and obtained velocity measurements. A compilation of all spectroscopically examined objects in DF2 can be found in Table \ref{tab:DF2}.

We include one additional GC (GC-80) in the sample and provide a clarification. 
GC-80 was observed with LRIS \citep{VanDokkum2018Nature} but was miscategorized as a background galaxy due to a bright OII emission line.  Subsequent spectroscopic follow-up by \citet{Emsellem2019} revealed that GC-80 (also called T3) is indeed a GC associated with DF2. The OII line possibly originated from a chance superposition of a galaxy in the background of this GC. A closer look at the HST image of this object reveals some faint extended structure around GC-80, which is consistent with a background galaxy.
In this paper, GC-80 is included in the confirmed GC sample. 

Since there is no new spectroscopic data for DF2 in this work, we simply take the locations and existing line-of-sight velocities from \citet{VanDokkum2018Nature} and \citet{Emsellem2019} and use them to label GCs and confirmed background galaxies. The photometric properties of the confirmed GCs are then used for the photometric measurements in Section \ref{sec:phot}.

\subsection{Keck/LRIS Observations of DF4}

Seven GCs have previously been identified in DF4 by \citet{VanDokkum2019}. We match the location of the seven GCs from the literature to the new photometric catalog and label them as confirmed GCs.

In addition to existing data, we obtained spectra of 12 new GC candidates using LRIS on the Keck I telescope. These candidates were selected from new deep HST imaging (see Section \ref{sec:phot} and \citet{Danieli2020}) which extends the previous footprint. \citet{VanDokkum2019} already observed all targets with $\langle F606W-F814W\rangle \pm 2\sigma$ and $V < 23$, where $\langle F606W-F814W\rangle$ and $\sigma$ are the mean and the standard deviation of the confirmed GC colors. To select new targets, we broadened the color range to $\langle F606W-F814W\rangle \pm 4\sigma$,
and included targets as faint as 24.5 mag, going deeper than the previous spectroscopic completeness limit of 23 mag. However, the new observations are not complete at 24.5 mag, due to geometrical limitations of the multi-object slit mask. Priority was given to bright targets and targets closer to the center of DF4. 

The new GC candidates were observed in a single slit mask on 2019 November 24. The weather was fair and seeing was around $1 \arcsec$. The total exposure time is 18,100\,s, split over nine  1800\,s  exposures  and  one  1900\,s  exposure. The high signal-to-noise blue-side spectra are used to confirm membership to DF4. 
The blue-side LRIS data were obtained with the 300/5000 grism and 10 slits, providing a spectral resolution ranging from $\sigma_{instr} \approx 350\, \text{km s}^{-1}$ at $\lambda = 3800$ \AA \
to $\sigma_{instr} \approx 150\, \text{km s}^{-1}$
at $\lambda = 6600$ \AA. 

We use the custom pipeline described in \citet{VanDokkum2018Nature} for reducing LRIS blue-side spectra. 
Since LRIS has considerable instrument flexure, each exposure was reduced and calibrated independently.
For each slit, an empirical sky model is generated and projected to the spatial direction with the appropriate line tilts. 
The wavelength solution is obtained in two steps: a rough estimate from the arc lamp calibration frame, and then the final solution from fitting the sky lines in each slit. 
Cosmic rays are cleaned by comparing sets of five sky-subtracted images and removing deviations above $4 \sigma$.
After each slit has been individually reduced, spectra from different frames are aligned and combined.
The 1D spectra are extracted from the combined 2D spectra and normalized by a 6th-order polynomial fit over the entire wavelength range 3800 to 6700 \AA.

The eleven combined and normalized 1D spectra are visually inspected. The presence of gas emission lines indicates whether an object is a background galaxy instead of a GC. Among all eleven candidates, five objects are background galaxies and one object is a foreground star. Spectra of the remaining five GC candidates are shown in Figure \ref{fig:spec}. All five spectra show prominent CaII H and K lines indicating old stellar populations.
In Table \ref{tab:DF4}, the eleven candidates are objects without ``Other ID'' entries since they were identified in this work.

\subsection{Radial velocities in DF4}
The radial velocities of the five GC candidates are obtained using \texttt{pPXF}\footnote{https://pypi.org/project/ppxf/} \citep{Cappellari2017}. \texttt{pPXF} combines stellar
template spectra and convolves them with a range of line-of-sight velocities v
and velocity dispersions $\sigma$ to produce a model spectrum that best
fits the logarithmic-rebinned galaxy spectrum and its kinematics. 
In this study, we use the template spectra of \citet{Vazdekis2010}, which are based on the MILES stellar library of \citet{SanchezBlazquez2006}. The 156 template spectra in this library span a metallicity range from -1.71 to +0.22, and an age range from 1 to 17.78 Gyr.

For each GC candidate, we only fit the region of our spectra blue-ward of
$5200$\AA\ in order to avoid the region worst affected by sky residuals. 
The templates are smoothed to the approximate spectral resolution of 
$\sigma_{instr} \approx 150\, \text{km s}^{-1}$. The best-fit GC velocities are listed in Table \ref{tab:DF4}.  Within the error bars, the GC velocities all agree with the NGC1052-DF4 diffuse light velocity of $1444\,\text{km s}^{-1}$ \citep{VanDokkum2019}. 

Table \ref{tab:DF4} contains a list of all point sources with spectroscopic data in DF4 to date. Where possible, the velocity measurements are listed as well. 
We list the confirmed GCs as well as background galaxy contaminants for future reference.

\section{Photometric Observations} \label{sec:phot}
In order to measure the faint end of the luminosity function beyond spectroscopic limits, we need to identify GC candidates from photometric data.

\subsection{DF2}

The HST imaging data for DF2 is the same as the data presented in \citet{VanDokkum2018Nature}, consisting of one orbit each with the ACS in F606W and F814W.
The photometry catalog is generated by \texttt{SourceExtractor} \citep{Bertin1996} in dual image mode. We follow the procedure in \citet{VanDokkum2018enigma} to derive total magnitudes for a consistency check.

During the process, we corrected a small error in applying the aperture correction of \citet{Bohlin2016}.  \citet{VanDokkum2018enigma} inadvertently used the diameter of 5 pixels as the radius. Correcting the radius to 2.5 pixels systematically offsets the $F606W- F814W$ colors by around 0.03 mag. While this correction changes the color criteria we quote in the next section, it has little impact on the final shape of the luminosity function because the same shift is also applied to the background photometry.

\subsection{DF4}

Prior to 2019, the HST data of DF4 consisted of one orbit each in F814W and F606W \citep{Cohen2018}. Since then, there has been seven additional orbits in the F814W filter and three orbits in the F606W filter, bringing the total exposure time up to 16,760\,s in F814W and 8,240\,s in F606W. 
The combined image reduction is presented in \citet{Danieli2020}.

We use the same \texttt{SourceExtractor} settings for DF2 and DF4, and the photometric corrections are calculated in the same way. This ensures that the photometic catalogs generated for both galaxies are compatible.

\subsection{Photometric Selection}

\begin{figure*}[ht!]
\centering
\includegraphics[width=0.75\textwidth]{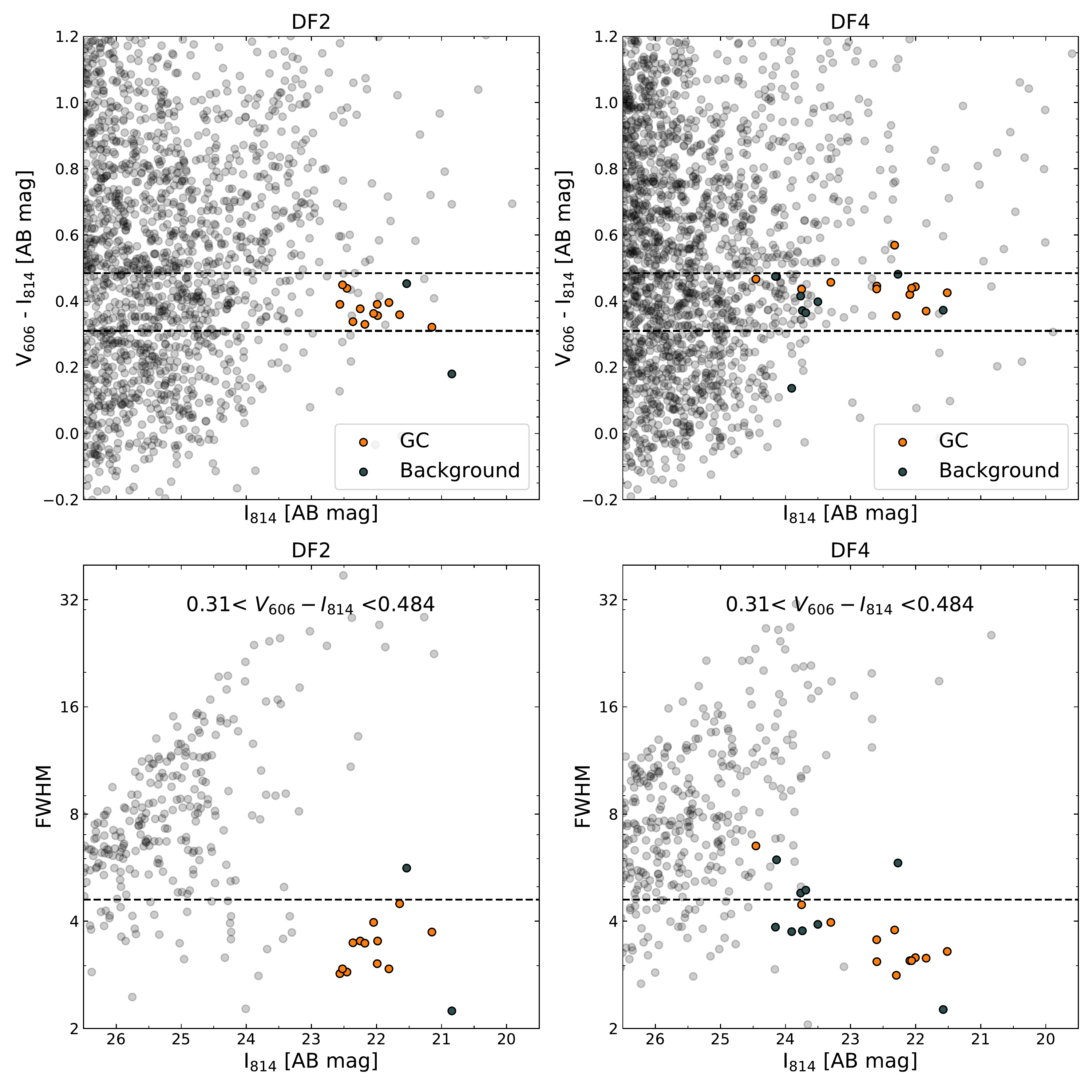}
\caption{Photometric selection criteria for globular clusters in DF2 and DF4. The left column is for NGC1052–DF2 and the right column is for DF4. Top row: color–magnitude relation of all objects in the HST image. Spectroscopically confirmed objects are marked with yellow and dark filled circles. Dashed lines delineate the $\pm 2 \sigma$ range of the colors of the confirmed clusters (excluding an outlier, GC-6571 with color of 0.54), which is displayed in text in the bottom panels. The bottom row shows the size–magnitude relation for all objects that satisfy this color criterion. Objects with FWHM below the dashed line are candidate GCs.  \label{fig:phot}}
\end{figure*}

\citet{VanDokkum2018enigma} selected GCs from HST photometry using a set of color and size criteria. Their selection cuts are calculated from the color and size distributions of the confirmed GC sample.
The color cut is $\langle F606W-F814W\rangle \pm 2\sigma = (0.282,0.438)$ and the size limit is $\langle FWHM\rangle + 2.5\sigma_{FWHM} = 4.7$ pixels. In our new analysis, the spectroscopically confirmed GCs in DF2 and DF4 are combined into one sample. As detailed in Section \ref{sec:spec}, one ``rediscovered’’ GC in DF2 and five new GCs in DF4 have been added to this sample. Based on their color and size distribution, we assume that the GC system in DF2 and DF4 are similar enough that the data can be combined. 

Besides using a combined sample, the other notable difference from \citet{VanDokkum2018enigma}
is that we excluded two outliers in DF4 for calculating the selection box. 
Figure \ref{fig:phot}  shows that two globular clusters in DF4 fall outside of the selection box in color and in size.
Both GCs are among the five new discoveries. The outlier in color is GC-6571 which is $4 \sigma$ away 
from the mean color; the size outlier is GC-97, which is more than $5 \sigma$ away from the other confirmed GCs. 
Although these two outliers are indeed GCs associated with DF4, we choose to exclude them temporarily. If these two were to be included, the selection box has to be significantly wider and thus allow more unrelated background sources to contaminate the final luminosity function and drastically increases the Poisson error in the background count. In order to reduce noise in the background subtraction step that follows, we exclude these two outliers and construct a narrow selection criterion.
While these two GCs are added back to the final count, we acknowledge that this may lead to undercounting possible GCs in the faint end of the final luminosity function. Rather than widening the selection box, we present a lower limit on the number of faint GCs in these two galaxies.

From a combined sample minus these two outliers, we obtain a color range $\langle F606W-F814W\rangle \pm 2\sigma = (0.31, 0.48)$ and size range FWHM $<\langle FWHM\rangle + 2.5\sigma_{FWHM} = 4.6$ pixels. 
Figure \ref{fig:phot} shows the distribution of point sources in DF2 (left panels) and DF4 (right panels). The top row are color-magnitude diagrams with the color range indicated as dashed lines. Only objects that satisfy this criterion are plotted in the bottom row. The same selection box, constructed from the combined GC sample, is applied to both DF2 and DF4.

\begin{figure*}[ht!]
\centering
\includegraphics[width=0.8\textwidth]{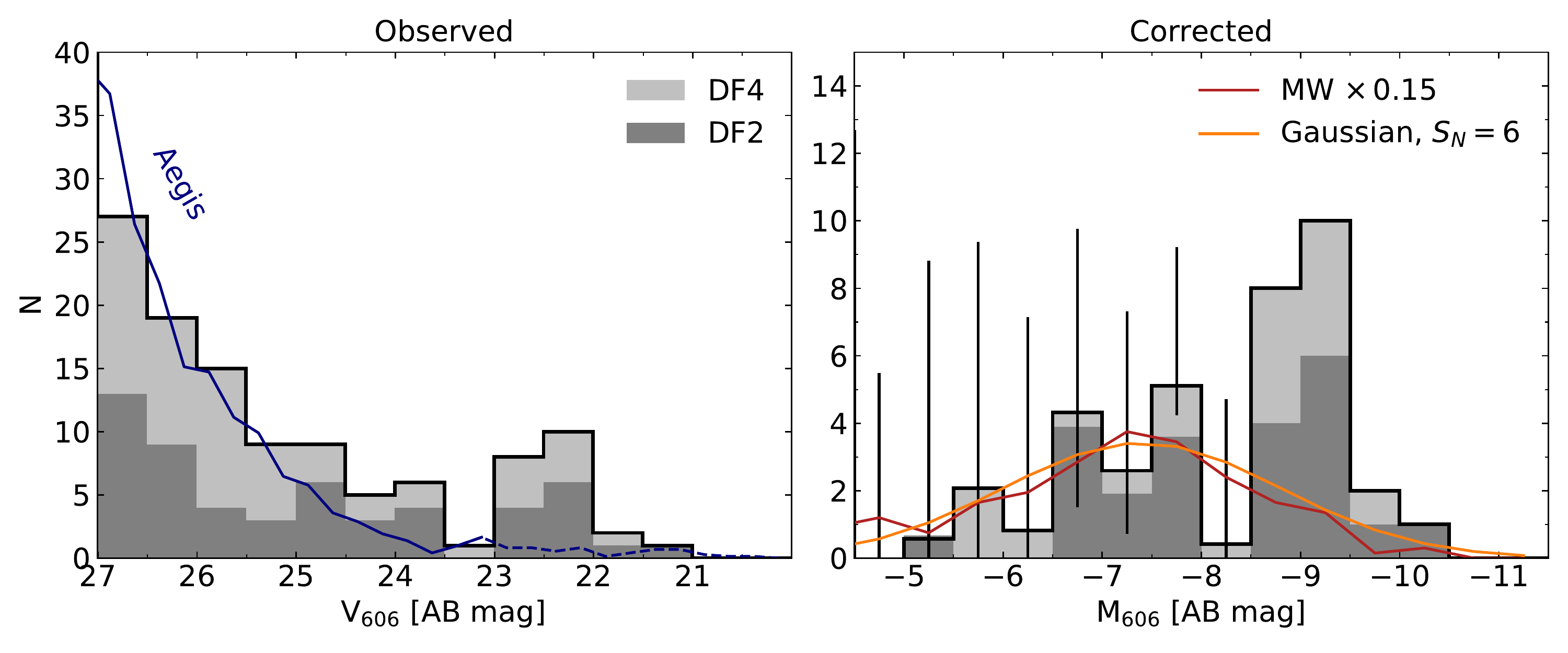}
\caption{Combined DF2 and DF4 luminosity function. Left panel: observed luminosity function, in apparent magnitude. The blue line shows objects in the 3d-HST/CANDELS blank field image that have the same colors and sizes as the observed GCs, which represents the expected number of unrelated background objects. Right panel: the background subtracted luminosity function, in absolute magnitude. Asymmetric error bars come from the 16th/84th percentiles of the exact Poisson distribution.
The luminosity functions of Milky Way GCs is overplotted in red, and a Gaussian luminosity function scaled by the stellar mass of DF2 and DF4 is plotted in orange. }\label{fig:lf_stacked}
\end{figure*}

\section{Results}
\subsection{Luminosity Function}

Background subtraction is necessary for F814W<23 mag because spectroscopic observations are incomplete below this magnitude. Some background sources are expected to fall inside our selection criteria. 
In the right-hand side panels of Figure \ref{fig:phot}, there is a mix of GCs and background sources with similar colors and sizes. 
The distribution of background sources is determined from the All-wavelength Extended Groth Strip International Survey (AEGIS) field from the 3D-HST data release \citep{Skelton2014}. 
The F814W and F606W AEGIS images are analyzed in the same way as the DF2 and DF4 data.

Figure \ref{fig:lf_stacked} shows the DF2/DF4 combined GCLF, before (left) and after (right) background subtraction.
For consistency with previous works, we display the luminosity function in $M_{F606W}$ magnitudes. 
Assuming a distance of 20 Mpc, the $M_{F606W}$ magnitudes are determined by:
$M_{F606W} = F814W+ (F606W -  F814W) -31.50$.
The left panel shows the stacked distribution of all sources with GC-like colors and sizes in the DF2 and DF4 fields, with the curve showing the expected number of contaminants.
The right panel shows the corrected distribution after subtracting the number of background sources in each bin where F814W$<23$ mag. The GCLF peak remains at $M_V \approx -9$ mag.
New data from the current work extends the faint-end constraints at $M_{F606W} > -8$, revealing a potential subpopulation of GCs at $M_{F606W} \approx -7.5$.
The main source of uncertainty in the faint end of the luminosity function comes from the number of background sources. The asymmetric error bars in the right panel are from the 16th/84th percentiles of the corresponding exact Poisson distribution. Error bars are only shown where the background is subtracted statistically.

For comparison, the GCLF of the Milky Way is overplotted in red (data from \citet{Harris2001}).
The Gaussian GCLF expected for DF2/DF4 is overplotted in orange, with the distribution $N(\mu = -7.4, \sigma =1.2)$ and the normalization is set by assuming a GC specific frequency $S_N=6$ for each galaxy. The GC specific frequency is defined as $S_N=N_{GC} \times 10^{0.4\times(M_{v,gal} +15)}$ \citep{Harris_vandenBergh1981,Harris1991}. For DF2 and DF4, $M_{v,gal} \approx -15$ so that the number of GCs in each galaxy is $N_{GC} = S_N = 6$.



\subsection{Faint GC-like Objects}

\begin{figure*}[ht!]
\centering
\includegraphics[width=0.8\textwidth]{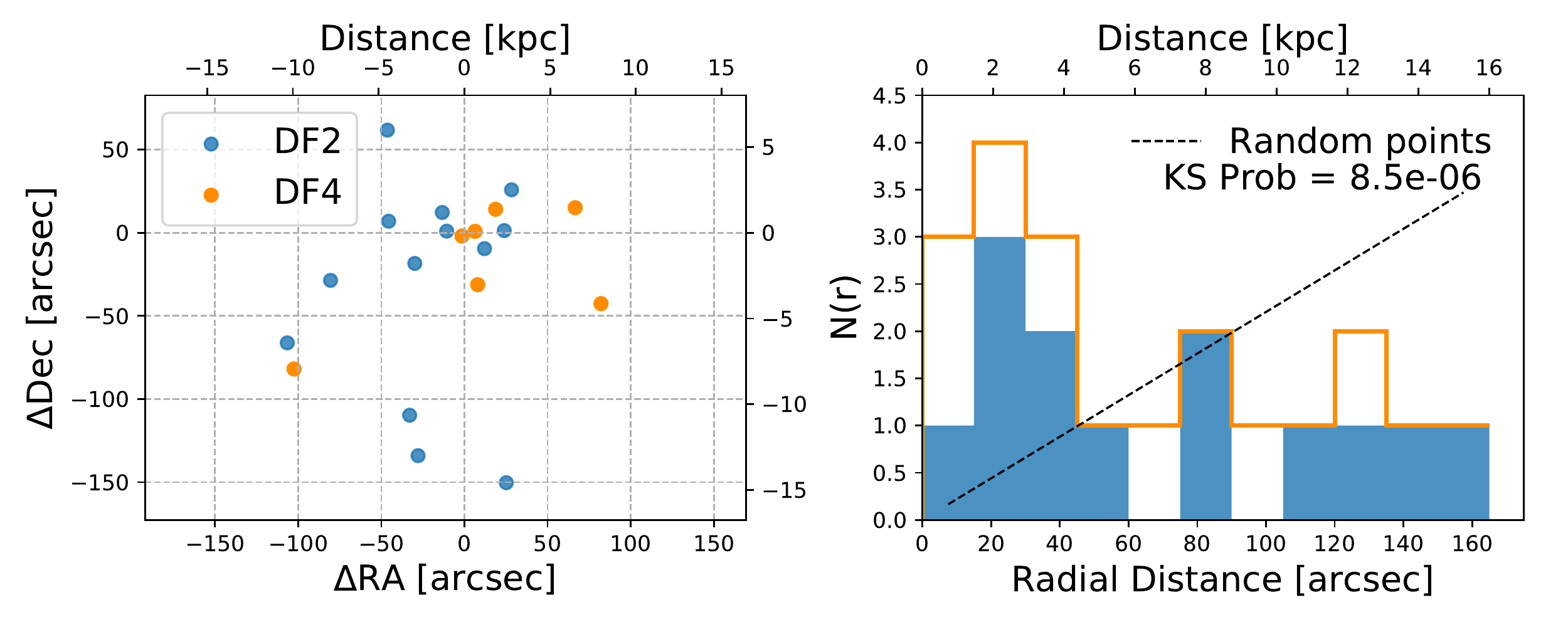}
\caption{Left panel: Spatial distribution of sources with $23 < F606W <25$ mag,
$0.31<F606W-F814W<0.48$, and FWHM $<4.6$, relative to the center of the galaxy. The angular separation is converted to physical separation assuming a distance of 20Mpc.
Right panel: Stacked radial distributions of GC candidates in DF2 and DF4. Dashed line is the expectation from a uniform distribution of sources. The uniform line is normalized to have the same area under the curve as the histogram. There is clearly an excess of GC candidates near the center, while the radial distribution is consistent with randomly distributed background sources at large radii. ``KS Prob'' refers to the significance level of the K-S test. A small probability means
more different distributions.
}\label{fig:faint_loc}
\end{figure*}

In the background-subtracted GCLF in Figure \ref{fig:lf_stacked}, there is a small peak at $M_V \approx -7.5$, coinciding with the near-universal GCLF peak of \citet{Harris1991}. If these faint sources are indeed GCs associated with DF2 and DF4, then this suggests a subpopulation of GCs with ``normal'' or expected luminosities.
In this section, we test the hypothesis that there is a population of faint (or ``normal'') GCs and compare their properties to those of the confirmed GCs.

Sources in both galaxies that contributed to the $M_V \approx -7.5$ bump in the GCLF are identified by the criteria $23 < F606W <25$ mag,
$0.31<F606W-F814W<0.48$, and FWHM $<4.6$ pixels. 
Approximately $75\%$ of these sources are expected to be unrelated objects (stars and background galaxies).
At this magnitude, individual sources are too faint for spectroscopic confirmation, but we can study the spatial distribution of these sources.
For all these sources in both DF2 and DF4, their positions relative to the galactic center is plotted in the left panel of Figure \ref{fig:faint_loc}. 
The a priori distribution of background objects is random while GCs should be clustered around the galaxy center.
Fig. \ref{fig:faint_loc} shows that the faint sources are spatially clustered around the center, which indicates that they contain real GCs.
For a quantitative test, the radial distance distribution of these sources is plotted in the right-hand panel of Figure \ref{fig:faint_loc}. The radial distance of sources in both galaxies are histogrammed in equal-width bins and stacked. 
If the distribution were random, the cumulative distribution of the radial distances should be $F(r)_{random} \propto r^2$ since the area covered in each bin increases as radius squared.
Thus, the probability distribution in each radial bin is $N(r)_{random} = \frac{dF}{dr} \propto r$. 
The expected number of sources in each bin from a uniformly random background is proportional to the distance and is normalized to the total number of observed sources.
This is the case at large radii, where the data is consistent with random distribution. However, there is clearly an excess of sources near the center of both galaxies. 

A Kolmogorov-Smirnov (KS) test is conducted to compare the cumulative distribution of all radial distances to the expected distribution from random sources.  A one-sample KS test of the 18 radial distances (corresponding to all points shown in the left panel of Figure \ref{fig:faint_loc}) against the random distribution yields a p-value of $8.5 \times 10^{-6}$, which means we can reject the hypothesis that the points are randomly distributed. 
From the spatial distribution of the faint GC-like sources, we subtract the expected distribution of random sources (i.e. $N_{observed} - N_{random}$) and find that 43\% of them are associated with DF2 and DF4. This is a higher fraction than the estimated 25\% from subtracting AEGIS blank field.

If the faint GCs form a different subpopulation, their sizes and colors may differ from the confirmed bright GC subpopulation.
However, we compared the \texttt{SourceExtractor} FWHM of the faint sources (mean = 3.74 pixels) and the confirmed GCs (mean = 3.53 pixels) and find no significant difference between the two samples. Both samples have mean color of 0.41, and a two-sample KS test shows $p = 0.82$. The faint sample and the confirmed sample are similar in FWHM and color.

\subsection{Gaussian Modeling}

Having established that the faint objects in the GCLF is associated with DF2 and DF4, we now investigate whether the faint bump is consistent with the expectation for ``normal'' low luminosity galaxies.

\begin{figure}[ht!]
\centering
\includegraphics[width=\columnwidth]{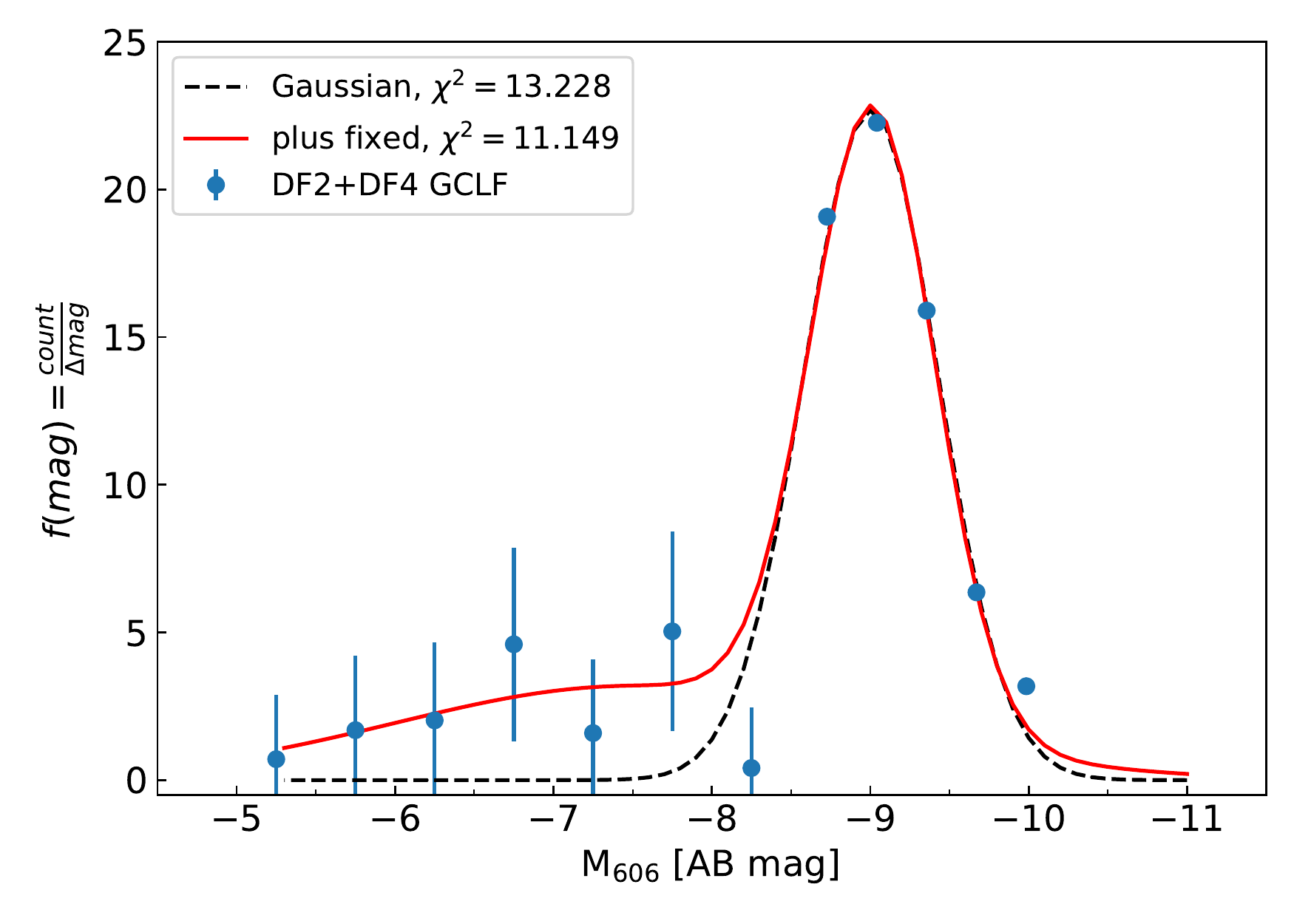}
\caption{The stacked DF2+DF4 globular cluster luminosity function in blue data points, where the confirmed portion ($M_{F606W} < -8.5$) have no error bars. Overplotted are two models: the black dashed line shows a single Gaussian fit, and the red line model has an additional fixed Gaussian component with amplitude of 12, center at $M_V = -7.5$, and sigma of 1.5. Both models have the same number of free parameters (D.O.F. = 9), and the same number of data points. The chi-square indicates that the two-component model is a better fit to the luminosity function. 
}\label{fig:twogaus}
\end{figure}

The near-universal GCLF is a Gaussian $N(\mu=-7.5,\sigma=1.5)$ normalized to the number of GCs in the galaxy \citep{Harris1991}. We use a GC specfic frequency \citep{Harris_vandenBergh1981} of 6, which translates to 12 GCs based on the luminosity of DF2 and DF4.
The GCLF is fitted with two Gaussian models:
the one-component model is a Gaussian: $A\times N(\mu, \sigma)$; the two-component model has an additional fixed Gaussian component: $A\times N(\mu, \sigma) + 12 \times N(-7.5,1.5)$.
We use the \texttt{lmfit} Python package to carry out non-linear least-squares curve fitting.
The models are fit to the binned and stacked DF2+DF4 GCLF using a standard Levenberg-Marquardt algorithm.
Spectroscopic information is complete on the GCs with $M_{F606W}< -8.5$. To better constrain the fits, the luminosity function in this range is rebinned into five equal width bins and the value of the luminosity function is obtained by dividing the count in each bin by the width of the bin. This luminosity function is shown in blue points in Figure \ref{fig:twogaus}, where the confirmed GCs have no error bars.

The error estimate used in the fitting procedure is based on the Poisson error on each data point. Since the data consists of small numbers, the Poisson error is not simply $\sqrt{N}$. Instead, the asymmetric upper and lower errors are determined from the exact Poisson error formula and the larger upper error is used for the least-squares fit.
Since the data can be understood as a sample from the underlying luminosity distribution, the errors should come from Poisson error of the model.
In this case, the error estimate is kept the same between two models to eliminate its effect on the chi-square statistic and ensure the comparison is fair. Thus the errors come from the count in each magnitude bin in the observed GCLF, and not from the underlying model.

The best-fit curves of both models are shown in Figure \ref{fig:twogaus}.
The two models have an equal number of free parameters and are fitted to the same data points. Both models have 9 degrees of freedom, which allows us to directly compare their chi-square statistics.
Neither model can be ruled out with more than 90\% confidence.
The chi-square of the one-component model is 13.228 (reduced chi-square = 1.47) and the chi-square of the two-component model is 11.149 (reduced chi-square = 1.24). Although bimodality in the observed GCLF cannot be statistically established, the two-component model does provide a better description of the data.

\subsection{Summary}

From a combined analysis of GCs in DF2 and DF4, we obtain two main results:
\begin{enumerate}
    \item the combined GCLF peaks at $M_V \approx -9$, as identified by \citet{VanDokkum2019}. 
    \item We also find a secondary peak at $M_V \approx -7.5$, coinciding with the near-universal GCLF peak \citep{Harris1991}.
\end{enumerate}

DF2 and DF4 each have twelve spectroscopically confirmed GCs.
The estimate of the total GC number is $18.5_{-4.42}^{+8.99}$ for DF2 and $18.6_{-4.92}^{+9.37}$ for DF4, calculated from the background-subtracted GCLF between $M_V = -5$ and $M_V=-11$.
The updated total number of GCs in both galaxies is $37 ^{+11.08}_{-6.54}$.

\subsection{Is this GCLF the expected Gaussian?}
Statistical tests show strong evidence against the null hypothesis that the combined GCLF originates from a Gaussian parent distribution of N(-7.5, 1.5).
\begin{itemize}
    \item The KS test statistic is 0.9999997, with a p-value of order $10^{-85}$. The discrepancy between the data and the expected distribution is significant.
    \item The Shapiro-Wilk test results in W = 0.8, p = 0.0072. This means we can reject the Gaussian hypothesis with 99\% confidence.
    \item The Anderson-Darling test shows that $A^2 = 1, p<0.01$. This shows that we can reject the null hypothesis with 99\% confidence, which is consistent with the other test results.
\end{itemize}
From the above tests, it seems clear that the combined GCLF deviates significantly from the near-universal GCLF. The results remain the same for DF2 data alone. These results are not sensitive to the assumed standard deviation of the near-universal GCLF.

\subsection{Can the new GCs belong to NGC1052?}

As evident from Figure \ref{fig:intro}, the spatial distribution of globular clusters around DF2 and DF4 are extended. 
This creates a challenge both in selecting possible GC targets and in establishing membership.

The DF2 radial velocity is around $1805 \pm 1.1 \text{km s}^{-1}$, far from the group mean velocity of $1438 \pm 25 \,\text{km s}^{-1}$. Thus, GCs associated with DF2 are clearly distinct from NGC1052 group. DF4 has a mean velocity ($1444 \pm 7.8 \,\text{km s}^{-1}$) close to the NGC1052 group velocity, and spectroscopy alone cannot distinguish between the two. Given that the DF4 GC candidates have velocities that associate them at least with the NGC1052 group, we investigate the possibility that they belong to NGC1052 instead of DF4. 

\citet{Forbes2001} characterized the surface density $\Sigma\, [\text{deg}^{-2}]$ of globular clusters in NGC1052. The radial distribution of GCs follows a power law: $\Sigma \propto r^{-2.08\pm 0.13}$, where $r$ is the angular separation in arcsecs. 
Based on this distribution and the distance from NGC1052 to DF4, only 0.3 GCs are expected in the \textit{HST} field (described in Section \ref{sec:phot}). 

In addition, the faint GCs are clustered around DF2 and DF4 (see Figure \ref{fig:faint_loc}), which makes it highly unlikely that they are contaminants from NGC1052. 

\section{Conclusions}

In this paper we have conducted an updated and combined analysis of the GCLF of DF2 and DF4. Given a confirmed distance of 20 Mpc and incorporating new HST and Keck data, we expanded the GC sample size and updated the selection criteria. The resulting GCLF shows two unexpected features: the peak is 1.5 mag brighter than the canonical peak location, and there is evidence for a smaller subpopulation of GCs at the canonical peak location.

With a uniform analysis of both galaxies, we find evidence for a subpopulation of GCs at the expected luminosity of $M_V = -7.5$, especially in DF2. 
The lack of spectroscopic confirmation at this magnitude makes it hard to pin down the number of faint GCs or where they are in the galaxy. This implies that the GCs of DF2 may not be a homogeneous group, even though they are all metal-poor.

\begin{figure}[ht!]
\centering
\includegraphics[width=\columnwidth]{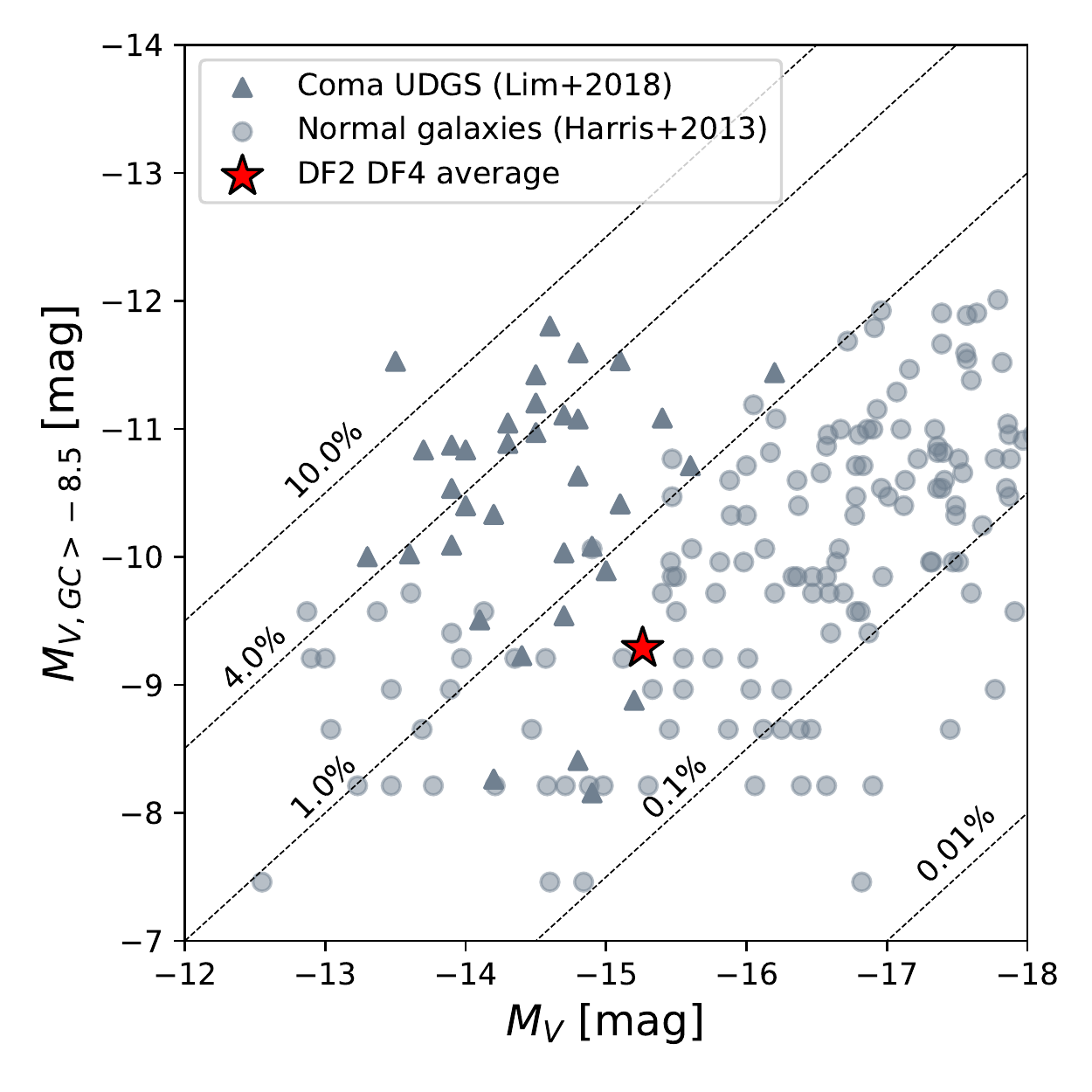}
\caption{The total absolute magnitude of GCs in NGC1052-DF2 and DF4 fainter than $-8.5$, $M_{V,GC >-8.5}$ vs. host galaxy total absolute magnitude $M_V$, shown with the red star. Grey circles are averages $M_{V,GC >-8.5}$ for normal galaxies derived from the literature compilation of \citet{Harris2013} and grey triangles are Coma UDGs from \citet{Lim2018}. Dashed lines represent globular cluster system to host galaxy absolute magnitude ratios. The faint GC population in NGC1052-DF2 and DF4 contribute to less than $1\%$ of the host galaxy absolute magnitude, which is common in both Coma UDGs and normal galaxies.}
\label{fig:discussion}
\end{figure}

Figure \ref{fig:discussion} shows the total absolute magnitude of GCs fainter than $-8.5$ in DF2 and DF4 compared to the host galaxy absolute magnitude. The faint GCs contribute less than $1\%$ of the total magnitude, similar to many normal galaxies and Coma UDGs. 
The total GC luminosities for DF2 and DF4 are calculated by a direct sum of the background-subtracted GCLF. For the galaxies in \citet{Lim2018} and \citet{Harris2013}, we assume a Normal GCLF with center of $-7.5$ and standard deviation of $1$, only considering the faint end above $-8.5$, and calculate the average GC absolute magnitude. Multiplied by the total number of GC from literature, this yields the total GC magnitude $M_{V,GC >-8.5}$.
The faint GCs in DF2 and DF4 are typical both in their luminosity function (which peaks at the canonical $-7.5$) and in the total luminosity (shown in Fig. \ref{fig:discussion}). What makes the GC system in DF2 and DF4 spectacular is the bright GCs around $M_V \approx -9$.


The number of GCs in DF2 and DF4 does not scale with DM halo mass, neither for the normal nor the over-massive population. This suggests that the observed correlation between $N_{GC}$ and $M_{halo}$ in ``normal'' galaxies is not fundamental, as found by \citet{Trenti2015, Boylan-Kolchin2017,Madau2020}. The correlation could emerge from a more fundamental correlation with the baryons \citep[e.g.,][]{Kruijssen2015,Bastian2020}. Given the low dark matter content of DF2 and DF4, they provide a good test for these two interpretations. 


In contrast to previous findings on the near-universal GCLF \citep{Harris1991,Richtler2003}, the observed GCLF in these two galaxies are offset from the canonical peak location. Although the  intrinsic scatter of the GCLF is claimed to be 0.1-0.2 mag \citep{Rejkuba2012GCLF}, we seem to have found an outlier that is 1.5 mag brighter. 
A possible origin of luminous GCs in DF2 and DF4 is that their environment might be highly conducive to GC-GC merging, but new simulations show that to be unlikely \citep{DuttaChowdhury2020}.
Galaxy-galaxy interactions can also trigger massive star cluster formation, or perturb the orbits of pre-existing GCs \citep{Leigh2020}. In this formation scenario, the luminous population could have formed in a merger of two galaxies, each of which already had a normal GC population. \citet{Trujillogomez2020} predicts that the luminous GCs formed in a gas-rich merger while the low-mass GCs were accreted from the progenitors, producing a bimodal GCLF. This scenario would be supported if the metallicities of low-mass GCs are clearly distinct from those of the massive GCs.
Any proposed formation scenario for NGC1052-DF2 and NGC1052-DF4 should account for both the extremely luminous GCs and the normal GCs.

\acknowledgments
We thank Diederik Kruijssen for insightful discussions on the project.
Support from STScI grant HST GO-15695 is gratefully acknowledged.
ZS is supported by the Gruber Science Fellowship.
S.D. is supported by NASA through Hubble Fellowship grant HST-HF2-51454.001-A awarded by the Space Telescope Science Institute, which is operated by the Association of Universities for Research in Astronomy, Incorporated, under NASA contract NAS5-26555.

%

\vspace{5mm}
\facilities{HST(ACS), Keck(LRIS)}


\software{astropy \citep{astropy2013,astropy2018},  
          numpy \citep{numpy2011},
          scipy \citep{SciPy2020-NMeth},
          Pandas \citep{pandas2020_reback,mckinney-proc-scipy-2010},
          matplotlib \citep{matplotlib2007},
          pPXF \citep{Cappellari2017},
          SourceExtractor \citep{Bertin1996}
          }


\appendix

\section{Alternative Photometric selection} \label{sec:prob}
\begin{figure*}[ht]
\centering
\includegraphics[width=0.75\textwidth]{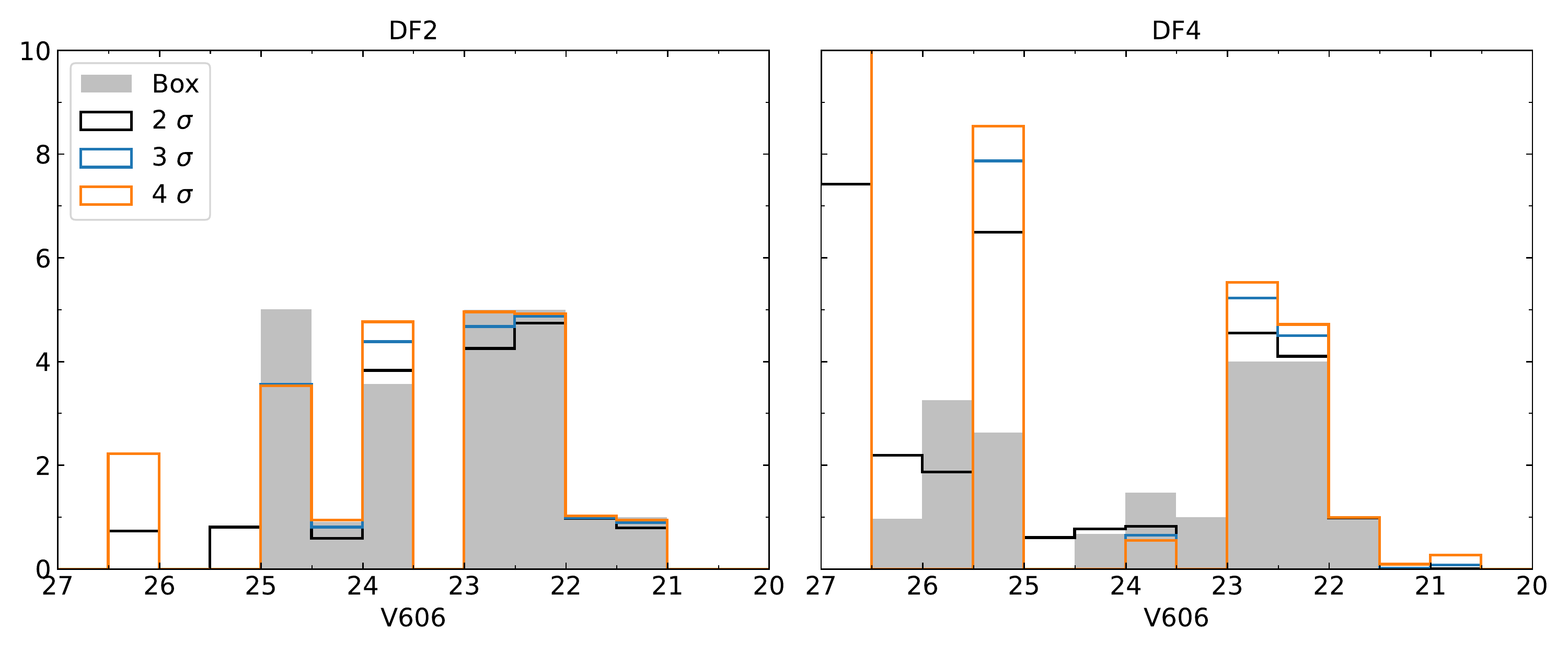}
\caption{Comparison between Gaussian probability selection and box criteria selection of GC candidates. }\label{fig:gaussian}
\end{figure*}

In this section we explore a probabilistic selection of GCs from the color-magnitude parameter space without placing any cuts on the data.
We rely on the color and size distribution of known GCs vs. known background sources to select possible GCs in the photometry catalog. Because the size distribution is unknown, we place a cut to only include sources below a certain size.
However, because the color distribution of GCs should be Gaussian, we can use a Gaussian probability associated with each point source to estimate the luminosity function. 
The Gaussian is determined by the confirmed GC sample and applied to all sources in the photometry catalog. Then, the background is estimated in the same way and subtracted off from the total count.
The statistical selection has the following steps.
\begin{enumerate}
    \item We select point sources from the photometry catalog below a size limit. The selection criteria is based on the spectroscopically confirmed GCs: the size limit is $\langle FWHM\rangle + 4\sigma_{FWHM}$. The criteria is intentionally broad so as not to miss any potential GCs.
    \item Each point source is assigned a weight based on a Gaussian probability distribution function: $$f(x) = \exp{ \left[ - \frac{(x-\mu)^2}{2\sigma^2} \right]}$$ where $\mu = \langle F606W-F814W\rangle $ and $\sigma = [2,3,4] \,\sigma_{V-I}$ in the confirmed sample.
    
    \item The sum of the weights for all sources in a magnitude bin is ``observed number of GCs'' in this bin.
    
    \item The above steps are repeated for the AEGIS catalog of known background sources, and subtracted from the ``observed'' luminosity function.
\end{enumerate}
As seen in Figure \ref{fig:gaussian}, the two methods yield very similar results for DF2, and the choice of 2, 3, or 4 $\sigma$ leads to minor differences in the shape of the LF.
Compared to a box criteria (e.g. in \citet{VanDokkum2019}), the advantage of the probability method is that it places no cuts on color.
Using the box criteria, a source has a weight of 0 if it falls outside the box and a weight of 1 if it is inside the box.
The Gaussian weights will be close to 0 if the color is far from the mean, and close to 1 if the color is close to the mean GC color.
The choice of the Gaussian (specfically the spread of $[2,3,4] \sigma$) is somewhat arbitrary, but does not impact the overall shape of the final luminosity function.

The relative disadvantage of the probability method is that background subtraction becomes ineffective at faint magnitudes (around 27 mag for DF2, and around 25 mag for DF4). It would be highly unlikely for DF4 to have more than 10 GC candidates that are five magnitudes fainter than the GCLF turnover point. The HST images of DF4 reach deeper than images of DF2, and the contaminants are not adequately subtracted using the probability method.

Since the probability methods does not provide a distinctive improvement upon the box criteria results, we presented this comparison in the appendix and did not use it as the main analysis tool. We chose the box criteria because it is more straightforward and the background subtraction process was cleaner.

\bibliography{paper.bib}{}
\bibliographystyle{aasjournal}

\end{document}